\documentclass[prb,twocolumn,superscriptaddress,amsmath,amssymb,showpacs]{revtex4}
\usepackage{graphicx}% Include figure files
\usepackage{dcolumn}% Align table columns on decimal point
\usepackage{bm}% bold math
\usepackage{txfonts}

\begin{document}
\title{Hysteretic response of electron-nuclear spin system in single InAlAs quantum dots:\par Excitation power and polarization dependences}

\author{R.\ Kaji}
	\affiliation{Department of Applied Physics, Hokkaido University, N13 W8, Kitaku, Sapporo 060-8628, Japan}
	%\email{r-kaji@eng.hokudai.ac.jp}

\author{S.\ Adachi}
	\affiliation{Department of Applied Physics, Hokkaido University, N13 W8, Kitaku, Sapporo 060-8628, Japan}
	\affiliation{CREST, Japan Science and Technology Agency, Kawaguchi 332-0012, Japan}
	\email{adachi-s@eng.hokudai.ac.jp}
	
\author{H.\ Sasakura}
	\affiliation{CREST, Japan Science and Technology Agency, Kawaguchi 332-0012, Japan}
    
\author{S.\ Muto}
	\affiliation{Department of Applied Physics, Hokkaido University, N13 W8, Kitaku, Sapporo 060-8628, Japan}
    \affiliation{CREST, Japan Science and Technology Agency, Kawaguchi 332-0012, Japan}

\date{\today}% It is always \today, today,

%%%%%%%%%%%%%%%%%%%%%%%%%%%%%%%%%%%%%%
\begin{abstract}
We report the hysteresis of optically-pumped nuclear spin polarization and the degree of circular polarization of photoluminescence on the excitation power and electron spin polarization in single InAlAs quantum dots. 
By increasing (or decreasing) the excitation power at a particular excitation polarization, an abrupt rise (or drop) and a clear hysteretic behavior were observed in the Overhauser shift of the photoluminescence of the exciton and exciton complexes from the same single quantum dot under an external magnetic field of 5 T. However, the degree of circular polarization shows different behaviors between a positively charged exciton and a neutral exciton or biexciton; further, only positively charged exciton exhibits the precisely synchronized change and hysteretic behavior. 
It is suggested that the electron spin distribution is affected by the flip-flop of electron-nuclear spins. Further, the hysteresis is observed as a function of the degree of circular polarization of the excitation light and its dependence on the excitation power is studied.
The saturation of the Overhauser shift after the abrupt rise indicates the almost complete cancellation of the external magnetic field by the nuclear field created within the width that is decided by the correlation time between the electron and the nuclei spin system. 
\end{abstract}
%%%%%%%%%%%%%%%%%%%%%%%%%%%%%%%%%%%%%%
\pacs{73.21.La, 78.67.Hc, 71.35.Pq,71.70.Jp}% PACS, the Physics and Astronomy
%\keywords{Zeeman splitting, exciton, quantum dots, hyperfine interaction}
%Use showkeys class option if keyword
\maketitle

%\section{Introduction} %%%%%%%%%%%%%%%%%%%%%%%%%%%%%%%%%%%%%%%%%%%%%%%%%%%%%%%%%%%%%%%%%%%%%%%%%%%%%%%%%
Recently, research on electron-nuclear spin interaction has been revived in view of its applications. This is because semiconductor quantum dots (QDs)
enhance the electron-nuclear spin interaction (hyperfine interaction) due to their strong 3D confinement of the electron wavefunction, and the enhanced interaction gives the possibility of aligning nuclear spins in one direction up to several tens \% in a single QD through the optical pumping. 
In fact, a large rate of nuclear spin polarization and the resultant large effective nuclear field up to several tesla were observed recently in interface GaAs QDs~\cite{Gammon01,Bracker05}, self-assembled InAlAs QDs~\cite{Yokoi05,Yokoi05b} and InGaAs QDs~\cite{Braun06,Tartakovskii07,Maletinsky07}.
Because of the ultralong coherence, nuclear spin is expected to contribute to applications such as a long-lived quantum memory at the nuclear level~\cite{Taylor03} and qubit conversion by using the nuclear field~\cite{Muto05}. 
Beyond such potential applications for quantum information processing, nuclear magnetic ordering and optically induced ferromagnetic ordering of spin systems are of surpassing interest in fundamental physics.
Therefore, the control of nuclear spins in nanostructures has both fundamental as well as  practical significance.  

In this study, we investigated the optical pumping of nuclear spin polarizations in a single self-assembled InAlAs QD. An abrupt rise in and the hysteresis of the Overhauser shift 
in addition to the degree of circular polarization (DCP) in the photoluminescence (PL) of positively charged excitons were clearly observed in the excitation power and excitation polarization dependences. Additionally, with the aid of this abrupt change, the sign of the electron $g$-factors in the $z$ direction is determined.

%\section{Sample and Experimental setup} %%%%%%%%%%%%%%%%%%%%%%%%%%%%%%%%%%%%%%%%%%%%%%%%%%%%%%%%%%%%%%%%%%%%
The QD sample was grown on a (100) GaAs substrate by molecular-beam epitaxy in the Stranski-Krastanow growth mode. The sample has an In$_{0.75}$Al$_{0.25}$As QD layer embedded in Al$_{0.3}$Ga$_{0.7}$As barrier layers~\cite{Yokoi05}.
For the single QD spectroscopy, small mesa structures with a typical top lateral size of $\sim$150 nm were fabricated. A cw-Ti:sapphire laser beam travelling along the QD growth direction was focused on the sample surface by a microscope objective. 
Time-integrated PL was measured at 5 K under the magnetic field up to 5 T in Faraday geometry.  The PL from QDs was dispersed by a triple-grating spectrometer, and it was detected with a liquid-nitrogen-cooled Si CCD camera.
The system resolution was found to be 12 $\mu$eV and the spectral resolution that determines the resonant peak energies was observed to be less than 5 $\mu$eV by using the spectral fitting. The typical measurement time of the CCD camera to obtain one spectrum with a high signal-to-noise ratio was 1 s. 

\begin{figure}[t]
  \begin{center}
    \includegraphics[width=220pt]{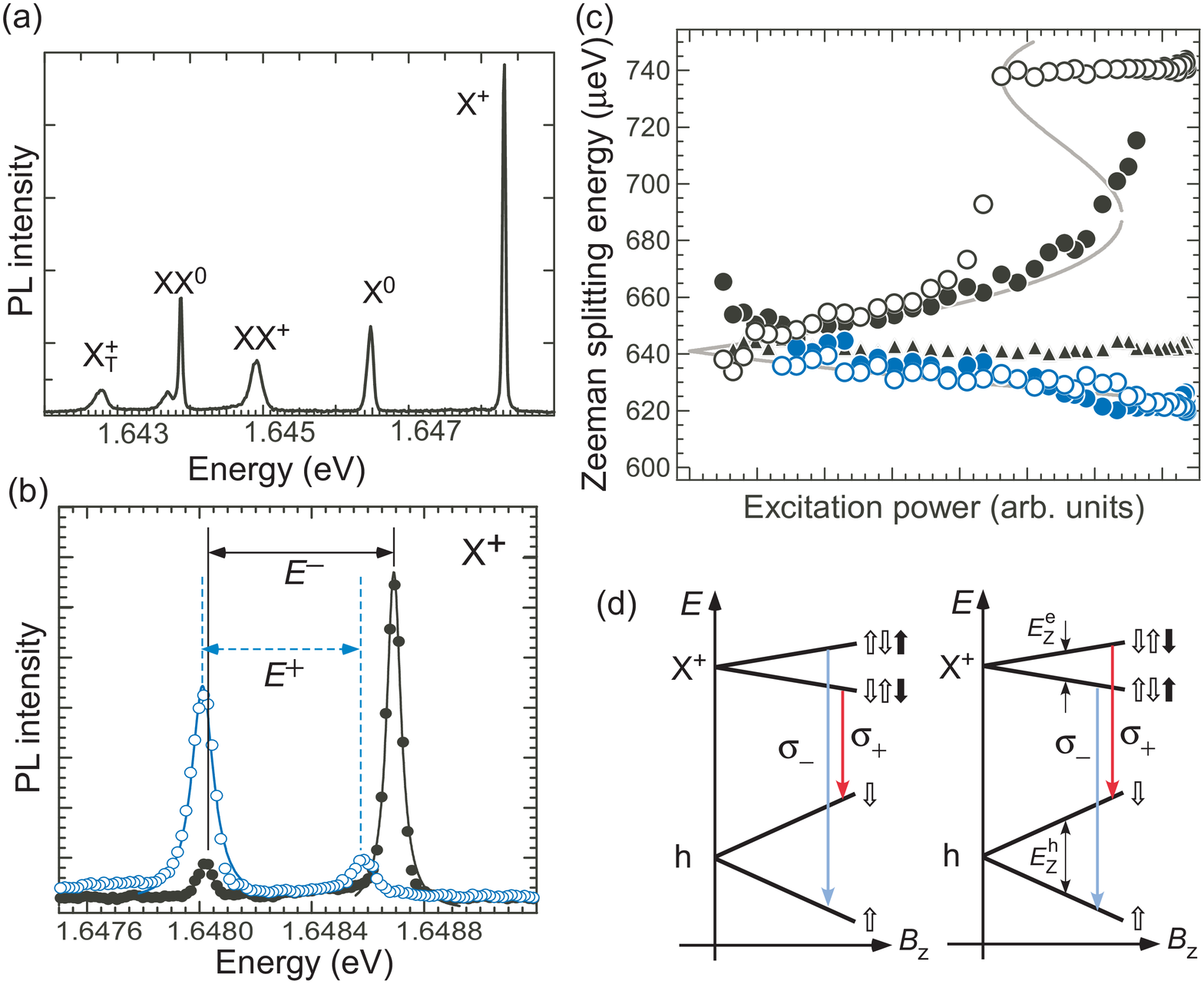}
  \end{center}
  \caption{(a) PL spectra of a single QD at 0 T (5 K). (b) PL spectra of $X^{+}$ at 5 T (5 K) for $\sigma_{+}$ (open circles) and $\sigma_{-}$ (solid circles) excitations. (c) Excitation-power dependence of $E^{+}$, $E^{-}$ (circles), and $E^{\rm L}$ (triangles) measured in the direction of increasing (solid marks) and decreasing excitation power (open marks) at 5 T. (d) Zeeman splitting in Faraday geometry and PL transitions of $X^+$ for $g_{\rm z}^{\rm e} \cdot g_{\rm z}^{\rm h} >0$ (left) and $g_{\rm z}^{\rm e} \cdot g_{\rm z}^{\rm h} <0$ (right) assuming $|g_{\rm z}^{\rm e} |<|g_{\rm z}^{\rm h}|$. The solid (open) up arrows and down arrows represent electrons (holes) with the total angular momentum $J_{e,z}^c \ (J_{e,z}^v)=+\frac{1}{2} \ \left(+\frac{3}{2}\right)$ and $J_{e,z}^c\ (J_{e,z}^v)=-\frac{1}{2} \ \left(-\frac{3}{2}\right)$.
  }\label{Fig1}
\end{figure}

%\section{Results~and~discussion} %%%%%%%%%%%%%%%%%%%%%%%%%%%%%%%%%%%%%%%%%%%%%%%%%%%%%%%%%%%%%%%%%%%%
The PL spectra of a target single InAlAs QD at 0 T is shown in Fig.~\ref{Fig1} (a). The spectra was obtained for the wetting layer (WL) excitation ($\sim$730 nm) by using depolarized light. The figure shows almost all the emission lines from an isolated QD for the WL excitation with the moderate power.  
Through various measurements for the assignment of the PL spectra~\cite{Kumano06}, we conclude that the PL lines in the figure originate from the same single QD and can be attributed to $X^{+}_{T}$ (triplet state of a positively charged exciton), $XX^{0}$ (neutral biexciton), $XX^{+}$ (positively charged biexciton), $X^{0}$ (neutral exciton), and $X^{+}$ (singlet state of a positively charged exciton) from the low-energy side. The binding energies are +2.5 meV and $-$1.8 meV for $XX^0$ and $X^+$, respectively. Furthermore, $X^0$ has a bright exciton splitting of $33 \pm 5 \ \mu$eV due to the anisotropic exchange interaction (AEI).
Hereafter, we focus on the $X^{+}$ PL because the strongest PL is emitted in the case of WL excitation. $X^{+}$ consists of spin-paired holes and an electron whose spin is selectively created  by circular excitation. Since $X^{+}$ has no dark states and since the spin-flipped electron can radiatively recombine with a hole immediately, the cycle rate for the electron-nuclear spin flip-flop process under cw-excitation is limited only by the fast spin-flip rate of the single-hole state. 
On the other hand, in the case of $X^{0}$, both bright and dark excitons can contribute to create the nuclear spin polarization, and the rate of spin flip-flop is limited by the long lifetime of the dark state in the case of the WL excitation. From the abovementioned features, the DCP of the $X^+$ PL is assumed for probing the electron spin directly. Therefore, the change of electron spins may become a mirror image of the change of nuclear spins. This argument will be tested later.
Figure~\ref{Fig1}(b) is a typical single-QD PL spectra of $X^{+}$ at 5 T for $\sigma_{+(-)}$-polarized excitation. We define $E^{+}$, $E^{-}$ and $E^{\rm L}$ as the respective Zeeman splitting energies for $\sigma_{+}$, $\sigma_{-}$ and linearly polarized excitations, as shown in the figure ($E^{\rm L}$ is not shown for simplicity, see Fig.~\ref{Fig1}(c)). 
The difference $E^{+(-)}-E^{\rm L}$ is known as the Overhauser shift (OHS)~\cite{Overhauser53}, and it can be explained by considering the optical pumping of nuclear spin polarization and the resulting nuclear magnetic field. Since the total magnetic field experienced by $X^{+}$ is the sum of the external magnetic field $B_{\rm z}$ and nuclear field $B_{\rm N}$, and since $B_{\rm N}$ acts only on electrons due to the nonzero existence probability at a nucleus site, Zeeman splitting for circular excitation in Faraday geometry can be written as
$\displaystyle E^{+(-)}=g_{\rm z}^{\rm h} \mu_{B} B_{\rm z} +g_{\rm z}^{\rm e} \mu_{B} (B_{\rm z} \pm B_{\rm N}),
$
where $g_{\rm z}^{\rm h(e)}$ is the hole (electron) g-factor in the $z$-direction and $\mu_{B}$ is the Bohr magneton.   
The equation says that the OHSs $E^{+(-)}-E^{\rm L}$ should have opposite sign, but be the same quantity corresponding to $\pm B_{\rm N}$ if $\pm B_{\rm N}$ does not depend on the excitation light polarization. In fact, such a symmetric OHS ($|E^{+}-E^{\rm L}|=|E^{-}-E^{\rm L}|$) has been observed in previous studies~\cite{Gammon01,Yokoi05,Yokoi05b,Mukumoto07}.
However, in this single InAlAs QD, $E^{+} - E^{\rm L} \sim -20 \ \mu$eV and $E^{-} - E^{\rm L} \sim +100 \ \mu$eV. This means that $B_{\rm N}$ is different for $\sigma_{+}$ and $\sigma_{-}$ excitations. 

Figure~\ref{Fig1}(c) shows the excitation power dependence of $E^{+}$, $E^{-}$, and $E^{\rm L}$ for $X^{+}$ at 5 T.
The solid (open) marks indicate the Zeeman energies measured in the direction of increasing (decreasing) excitation power. While $E^{+}$ ($E^{\rm L}$) indicates the gradual change (no change) with an increasing or decreasing excitation power, $E^{-}$ shows the abrupt change and bistable behavior on the excitation power. Such bistable behavior is observed only for $E^{-}$ in the $B_{\rm z}$ range of 2$-$5 T.
Recently, the similar bistable behavior of the OHS was observed for $X^0$ in InGaAs/GaAs QDs for an  excitation power in the range 1$-$3 T ~\cite{Tartakovskii07}.  
While, in a recent study, an abrupt change was observed in a decreasing $E^{+}$, in the data presented here an abrupt change appears only in $E^{-}$. This difference is due to the sign of $g_{\rm z}^{\rm e}$ in the InAlAs QD ($g_{\rm z}^{\rm e} \sim -0.37$)~\cite{Comment2}, which is opposite to $g_{\rm z}^{\rm e}$ in an InGaAs QD ($g_{\rm z}^{\rm e} \sim +0.60$). 
As observed in Fig.~\ref{Fig1}(d), the Zeeman splitting of $X^{+}$ PL is given as $E_{\rm Z}^{\rm h}+E_{\rm Z}^{\rm e}$ for $g_{\rm z}^{\rm e} >0$ and $E_{\rm Z}^{\rm h}-E_{\rm Z}^{\rm e}$ for $g_{\rm z}^{\rm e} <0$, where $E_{\rm Z}^{\rm h(e)}$ is the Zeeman splitting energy of a hole (an electron). Therefore, increasing the Zeeman splitting of $E^{-}$ in Fig.~\ref{Fig1}(c) signifies a \textit{reduction} in the electronic Zeeman energy $E_{\rm Z}^{\rm e}$ of $X^{+}$ due to the compensation of $B_{\rm z}$ by $B_{\rm N}$. Hence, our observation and the recent findings are interpreted as follows; An abrupt change occurs only when $B_{\rm N}$ reduces the external field $B_{\rm z}$, and therefore the same physics is considered to underlie the observed bistable behavior.

The optical pumping of the nuclear spin polarization is described by the following rate equation~\cite{OptOrientation,Abragam61}:
\begin{equation}
\frac{{d\left\langle {I_{z} } \right\rangle }}{{dt}} = \frac{1}{{T_{\rm{NF}} }}\left[ Q \left( {\left\langle {S_{\rm z} } \right\rangle  - S_{0} } \right)-{\left\langle {I_{z} } \right\rangle} \right] - \frac{1}{{T_{\rm{ND}} }}\left\langle {I_{z} } \right\rangle, \label{eq1}
\end{equation}
where $\langle {I_{z} } \rangle$ and $\langle {S_{\rm z} } \rangle$ are the averaged nuclear and electron spin polarizations; $S_{0}$ the thermal electron spin polarization; $1/T_{\rm{NF}}$ and $1/T_{\rm{ND}}$ nuclear spin polarization and depolarization rates; and 
$Q \ (=\left[{I\left( {I + 1} \right)}\right]/\left[{S(S + 1)}\right])$ is the momentum conversion coefficient from an electron spin to a nuclear spin system. 
Nuclear spin diffusion is included in the depolarization term. 
Based on the general form of the spin-flip process in the precessional decoherence type~\cite{OptOrientation}, the spin transfer rate $1/T_{\rm{NF}}$ is given as follows by assuming the uniform electron wavefunction in a QD~\cite{Braun06,Tartakovskii07}:
\begin{equation}
\frac{1}{T_{\rm{NF}} }= \left[\frac{n_{e} \tau_{c}^2}{\tau_{R}} \left(\frac{A }{N \hbar} \right)^2\right] \bigg/ \left[1+ \left(\frac{\tau_{c}}{\hbar} \right)^2 \left(g_{\rm z}^{\rm e} \mu_{B} B_{\rm z} \pm A \left\langle {I_{z} } \right\rangle \right)^2 \right],
\label{eq2}
\end{equation}
where $A$, $N$, $n_{e}$, and $\tau_{R}$ are the hyperfine coupling constant, number of nuclei, electron density of $X^{+}$ in the QD, and lifetime of $X^{+}$, respectively. $\tau_{c}$ is the correlation time of the coupled electron-nuclei system with a broadening $\hbar/\tau_{c}$ that is basically decided by the electron-spin relaxation time of $X^{+}$. 
$g_{\rm z}^{\rm e} \mu_{B} B_{\rm z} \pm A \left\langle {I_{\rm z} } \right\rangle(=g_{\rm z}^{\rm e} \mu_{B} (B_{\rm z} \pm B_{\rm N}))$ represents $E^{\rm e}_{\rm Z}$ affected by $B_{\rm N}$.  
According to the Eq.~\ref{eq2}, the external field compensation by a nuclear field decreases $E_{\rm Z}^{\rm e}$ and accelerates the spin transfer from electrons to the nuclei at a particular polarization.
In this simple model, the coupled electron-nuclear spin system shows a static hysteresis loop in the relation between the OHS (=$A\langle I_z \rangle$) and three variable parameters: $n_e$ ($\propto$ excitation power), $\langle {S_{\rm z} } \rangle$ ($\propto$ excitation polarization), and $B_{\rm z}$. 

\begin{figure}[t]
  \begin{center}
    \includegraphics[width=200pt]{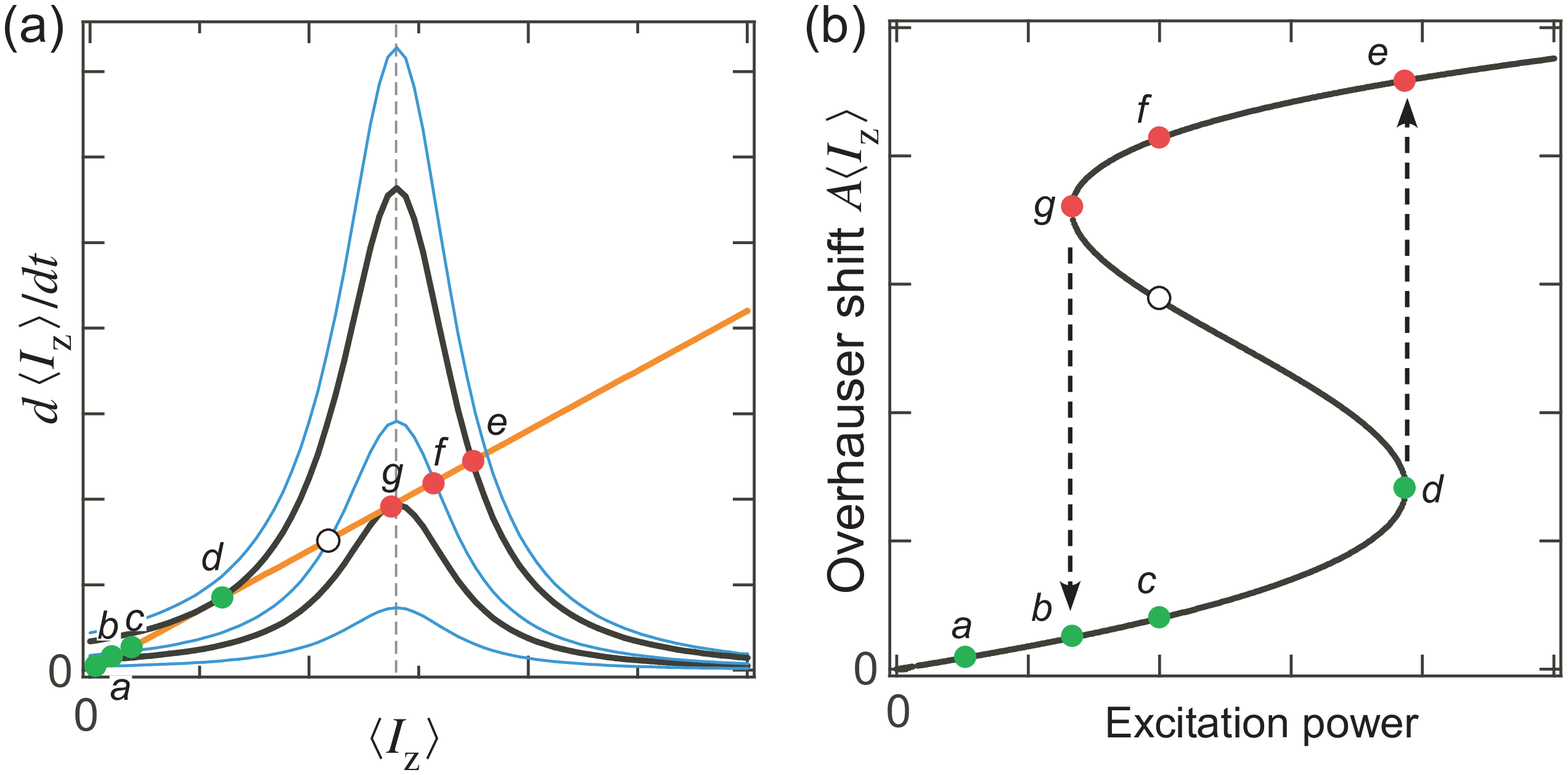}
  \end{center}
  \caption{(a) Graphical representation of excitation power dependence of OHS. The steady-state nuclear polarization is determined by the balance between the polarization (Lorentzian shape) and  the depolarization (straight line) terms. The thick curves represent the polarization terms at the low and high threshold excitation power. (b) Bistable behavior of the steady-state $\left\langle {I_{z} } \right\rangle$. The states $a-g$ correspond to those in (a).}%{}内にタイトルを記入してください
  \label{Fig2}
\end{figure}

In the next figure, we consider the excitation power dependence of the OHS.
The steady-state $\left\langle {I_{z} } \right\rangle$ of Eq.~\ref{eq1} with the rate given by Eq.~\ref{eq2} along with a constant $1/T_{\rm{ND}}$ are expressed graphically by the intersecting points of both the polarization and depolarization terms. 
The Lorentzian-shaped polarization term increases up with $n_{\rm{e}}$. 
Starting from the point $a$, the intersection is unique and then, with \textit{increasing} the excitation power, the steady-state $\left\langle {I_{z} } \right\rangle$ follows the straight line of the depolarization term. Beyond point $b$, two new solutions appear; however, the system still remains in the low $\left\langle {I_{z} } \right\rangle$ state. At point $d$, this lower state disappears and the OHS jumps up to a new state $e$. For further increasing the excitation power, the state remains on the upper branch. 
The trajectory of the steady-state $\left\langle {I_{z} } \right\rangle$ is depicted as a function of $n_{\rm{e}}$ in Fig.~\ref{Fig2} (b). 
In the case of a decreasing excitation power, the state on the upper branch remains down at the lower threshold power where the upper state $g$ disappears and the system has to return to the lower state (point $b$). In the region between the low and high threshold power, there are two stable $\left\langle {I_{z} } \right\rangle$; the realization of one of them depends on history, i.e., on whether one comes from higher or lower excitation power. The intermediate branch is unstable (e.g., open circle in the figure) and if the system is prepared by some means on this branch, the slightest deviation from the unstable branch due to a fluctuation causes the system to move into a state on the upper or lower stable branch.
According to this model, the experimental data in Fig.~\ref{Fig1}(c) can be fitted with the solid gray curve by using the following parameters: $N=4 \times 10^4$, $\tau_{\rm c}$=18 ps, $T_{\rm ND}$=10 ms, $\tau_{R}$= 1 ns, $g_{\rm z}^{\rm e}=-0.37$, and $\overline{A}=52.6 \ \mu$eV, $\overline{I(I+1)}$=12.25, where $\overline{A}$ and $\overline{I(I+1)}$ are the weighted averages for an In$_{0.75}$Al$_{0.25}$As QD.

\begin{figure}[t]
  \begin{center}
    \includegraphics[width=220pt]{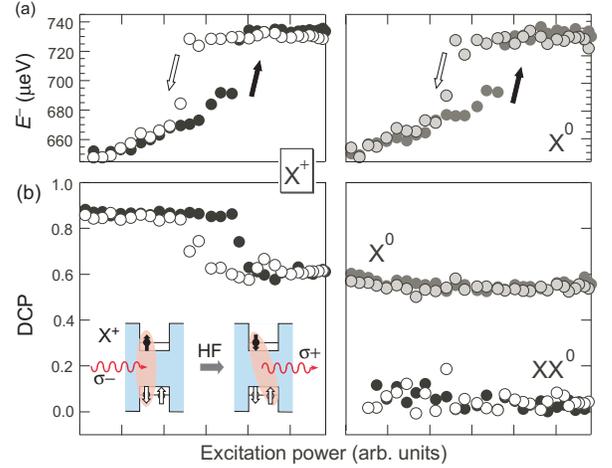}
  \end{center}
  \caption{(a) Observed $E^{-}$ as a function of the excitation power (5 K, 5 T). The black (white) arrows indicate the measurement direction of the power dependence plotted by the solid (open) circles. Hysteresis in $E^{-}$ were observed for $X^{+}$, $X^{0}$, and $XX^{0}$ PL ($E^{-}$ of $XX^{0}$ is not shown here, but it is almost the same as that of $X^{+}$ and $X^{0}$).
  (b) Observed hysteresis loop in the DCP for $X^{+}$ PL. Note that DCP is defined as $(I^{-}-I^{+})/(I^{-}+I^{+})$. This hysteresis can be observed for $X^{+}$ PL only (left), and it is not observed for $X^{0}$ and $XX^{0}$ PLs (right). Schematic of the $\sigma_{-}$-excitation of $X^{+}$ and the emission after the spin flip-flop process between an electron and a nucleus (inset). }\label{Fig3}
\end{figure}

Figure~\ref{Fig3} highlights the importance of $X^{+}$. 
$E^-$ for $X^{+}$ PL (left panel) and $X^{0}$ PL (right panel) shows the same bistable behavior and an abrupt change at exactly the same low and high threshold power. It is not surprising since the created $B_{\rm N}$ is effective for the entire target single QD. Therefore, all excitations such as $X^{+}$ and $X^{0}$ with single electron in the same QD should feel the same $B_{\rm N}$ if they have the same $g_{\rm z}^{\rm e}$. However, the DCP of the PL is quite different for $X^{+}$ and $X^{0}$, as observed in Fig.~\ref{Fig3}(b).  
Here, the DCP is defined as $(I^{-}-I^{+})/(I^{-}+I^{+})$, where $I^{+(-)}$ denotes the integrated PL intensity of the $\sigma_{+(-)}$-polarized spectrum.
Note that $XX^0$ PL is always a mirror image of $X^0$ PL in the Zeeman splitting energy and has zero DCP since $XX^0$ is not affected by $B_{\rm z}$ and $B_{\rm N}$ based on the spin-paired electrons and holes and has equal transition probabilities to the $X^0$ states with $J_z=\pm 1$.

While the DCP of $X^{+}$ shows an abrupt change and hysteresis synchronized to those for the OHS, the DCP of $X^{0}$ show no signature. These data clearly indicate that $X^+$ PL directly probes the electron spin.
In order to change the steady-state DCP following a change of OHS, 
an electron spin relaxation with a characteristic time comparable to the $X^+$ lifetime ($\sim$ 1 ns) is necessary when total magnetic field $B_{\rm z}-B_{\rm N} \sim 0$. 
Generally, the external longitudinal magnetic field can significantly suppress the electron spin relaxation in the "internal random magnetic field" (IRMF) that originates from the effective magnetic fields resulting from the hyperfine interaction, exchange interaction, and spin splitting of the conduction band~\cite{OptOrientation}. 
Therefore, when $B_{\rm z}-B_{\rm N}=0$, the above competing electron spin relaxation can emerge and reduce the steady-state DCP. In fact, the DCP is $\sim$0.35 and $\sim$0 for $X^+$ and $X^0$ for 0 T in this QD. 
The difference between the DCP for $X^0$ and $X^+$ is interpreted as follows. The DCP of $X^0$ is low because of the AEI. At 5 T, it is improved to 0.6, and with an increase in $B_{\rm N}$, it shows little change since the hole with a large $g_{\rm z}^{\rm h}$ ($\gg |g_{\rm z}^{\rm e}|$) still feels $B_{\rm z}$. On the other hand, the DCP of $X^+$ is high due to the absence of the AEI between electron and spin-paired holes. It is improved at 5 T due to the suppression of the electron spin relaxation by the IRMF; however it deteriorates after switching by the cancellation of $B_{\rm z}$ and the resulting revival of the suppressed relaxation. The origin of the IRMF in our QD has not been  identified yet. However, the spin splitting of the conduction band is not likely because it is not expected to cause spin relaxation in QDs. Therefore, the hyperfine interaction is the most likely origin of the IRMF. The electron spin relaxation due to the hyperfine interaction is nothing but the term described by Eq.~\ref{eq2}. Therefore, it suggests that the electron spin polarization is affected by the flip-flop of electron and nuclear spins. This is an unusual situation. Textbooks assume that the electron spin distribution is determined regardless of the electron-nuclear flip-flop and the only effect of the nucleus is due to the spin conserving term involving $B_{\rm N}$.

\begin{figure}[t]
  \begin{center}
    \includegraphics[width=240pt]{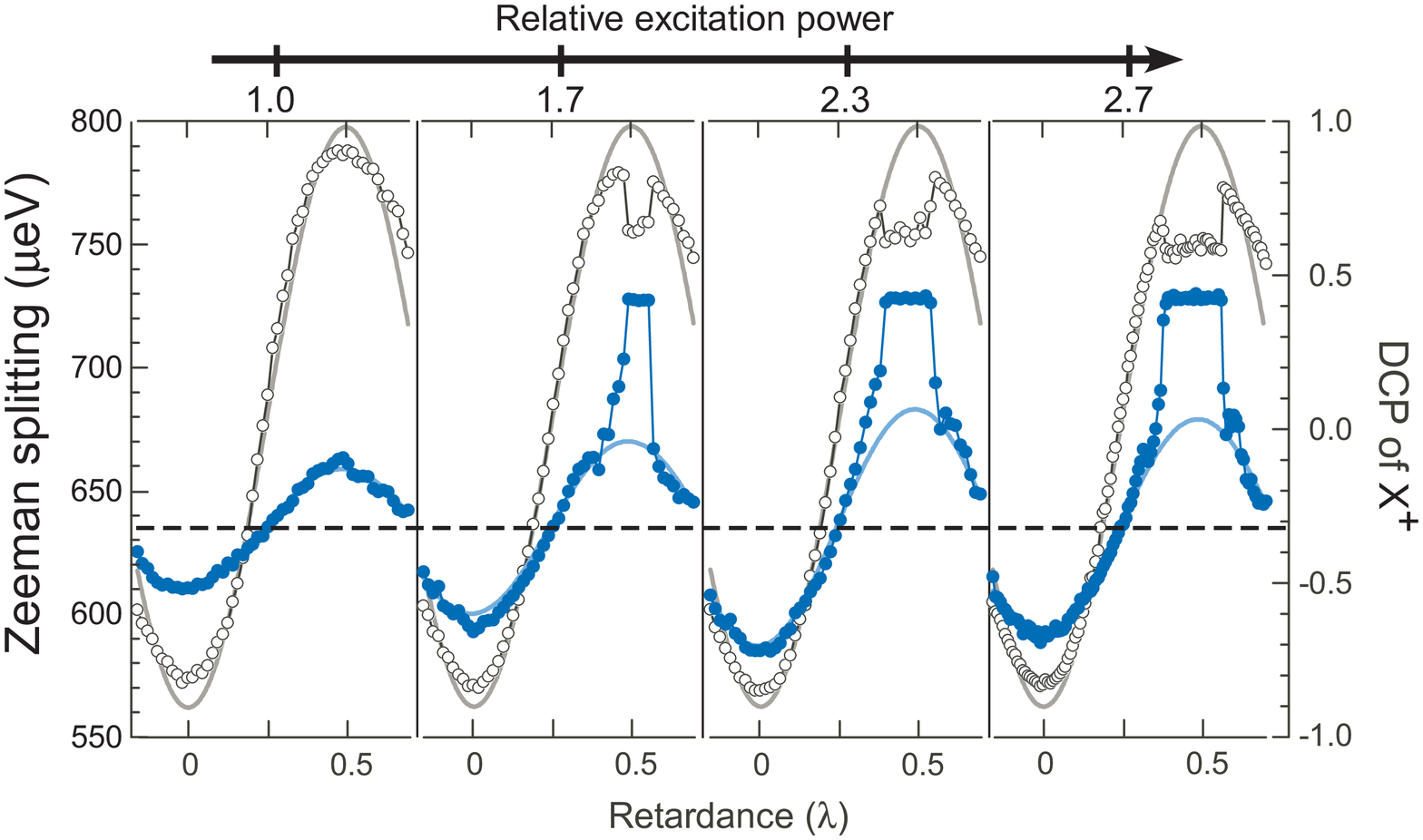}
  \end{center}
  \caption{Variation in Zeeman splitting energy that is affected by the nuclear field depending on the excitation polarizations (5 T, 5 K). The dashed line indicates the splitting $E_{\rm L}$. The corresponding reduction in the DCP (open circles) is also observed. The retardances $\lambda = 0$ and 0.5 correspond to the $\sigma_{+}$ and $\sigma_{-}$ excitations, respectively.}\label{Fig4}
\end{figure}
Finally, the excitation polarization dependence of Zeeman splitting and the DCP of $X^{+}$ at different excitation power are shown in Fig.~\ref{Fig4}. 
We note that the excitation polarization controls one of the most important parameters of the rate equation of Eq.~\ref{eq1}, i.e., $\langle S_{\rm z} \rangle$. It indirectly controls the quantity $\left[ Q \left( {\left\langle {S_{\rm z} } \right\rangle  - S_{0} } \right)-{\left\langle {I_{z} } \right\rangle} \right]$ in Eq.~\ref{eq1}, the zero of which determines the equilibrium value of $\langle I_z \rangle$ in the limit of negligible $1/T_{\rm ND}$. The value $\langle I_z \rangle =  Q \left( {\left\langle {S_{\rm z} } \right\rangle  - S_{0} } \right)$, in the limit of the lowest order of hyperfine constant, is the value given by the detailed balance argument in the textbook~\cite{Abragam61}. Therefore, the control of the light polarization presents an independent way of controlling the nuclear polarization switching in addition to the excitation power and the external magnetic field both of which control the quantity $1/T_{\rm NF}$. Unfortunately, in the present case, the control just resembles the excitation power dependence through the factor $n_{\rm e} \langle S_{\rm z} \rangle$ because $1/T_{\rm ND}$ is not small and $Q$ is large for an InAlAs QD, and therefore $\langle I_z \rangle$ can be neglected in the term $Q \left( {\left\langle {S_{\rm z} } \right\rangle  - S_{0} } \right) -\langle I_z \rangle $ and not in $\langle I_z \rangle /T_{\rm ND}$ .

In the leftmost panel, the Zeeman splitting and DCP follow the excitation polarization as ${\rm cos}(2 \pi \lambda- \pi)$ centering on $E_{\rm L}$ ($\sim 640 \ \mu$eV) and $S_0$ ($\sim 0.05$), respectively, where $\lambda$ is the retardance that is generated by rotating a quarter-wave plate in the excitation laser path. This is a case under the threshold excitation power~\cite{Yokoi05}. 
As exceeding the threshold excitation power, the deviation from $E_{\rm L}$ (i.e., OHS) increases and the energy shift changes abruptly by up to $\sim 100 \ \mu$eV around the $\sigma_{-}$ excitation ($\lambda \sim 0.5$). In addition, the synchronized reduction of the DCP occurs. With a further increase in the excitation power, the region of saturation of the OHS and DCP increases according to the product $n_{\rm e} \langle S_{\rm z} \rangle$ and becomes asymmetrical about $\lambda=0.5$, reflecting a bistable nature. It is worth pointing out that the maximum value of the OHS saturates as long as $n_{\rm e} \langle S_{\rm z} \rangle$ is above a given threshold. This means that the width of the Lorentzian-shaped polarization rate in Fig.~\ref{Fig2} is sufficiently small. Therefore, the complete cancellation of the external magnetic field occurs at the saturated OHS 
because the difference between $B_{\rm z}$ and $B_{\rm N}$ is within the narrow width of the Lorentzian that is decided by $\hbar / \tau_c$. The cancellation becomes closer to exact as $B_{\rm z}$ is large within the range in which the bistable behavior is observed since the saturated value occurs near the top of Lorentzian-shaped polarization rate.

%\section{SUMMARY}%%%%%%%%%%%%%%%%%%%%%%%%%%%%%%%%%%%%%%%%%%%%%%%%%%%%%%%%%%%%%%%%%%%%%%%%%%%%%%%%%%%%%
In summary, we investigated the optical pumping of nuclear spin polarizations in a single InAlAs QD and observed the clear bistable behavior of the OHS up to $\sim 100\ \mu$eV in the excitation power and polarization dependences at 5 T. The bistability can be seen only in the case of $\sigma_{-}$ excitation and the behavior is explained by the simple model in which the rate of spin flip-flop between an electron and a nucleus depends on the electronic Zeeman splitting that is affected by the nuclear field. We found that $X^+$ PL directly probes the electron spin and therefore the DCP synchronizes the change in the OHS. 
It is suggested that the electron spin distribution is affected by the flip-flop of electron-nuclear spins.
The saturated OHS shows the complete cancellation where the difference between the external magnetic field and the created nuclear field is within the narrow width of the Lorentzian function.

This work was financially supported by a Grant-in-Aid for Scientific Research from the Ministry of Education, Culture, Sports, Science, Japan.

\newpage


\begin{thebibliography}{99}
\bibitem{Comment2} It is worth pointing out that the abrupt change in the OHS can be used to obtain the magnitude and the sign of $g_{\rm z}^{\rm e}$. We obtained $g_{\rm z}^{\rm e}=-0.37 \pm 0.02$ and $g_{\rm z}^{\rm h}=2.54 \pm 0.01$for this InAlAs QD by measuring the abrupt change in the OHS; for this, $B_{\rm z}$ was scanned at several values of the excitation power.
\end{thebibliography}
\end{document}